Here the paper follows
\documentstyle[12pt]{article}

\def \nonu {\nonumber}

\def \c {\cite}

\begin{document}
\begin{titlepage}
\title {\bf QUARK HADRON PHASE TRANSITION AND HYBRID STARS }
\author{\bf Sanjay K. Ghosh, S. C. Phatak and Pradip K. Sahu \\
Institute of Physics, Bhubaneswar-751005, INDIA.}
\maketitle
\begin{abstract}
We investigate the properties of hybrid stars
consisting of quark matter in the core and hadron matter in outer
region. The hadronic and quark matter equations of state are
calculated by using nonlinear Walecka model and chiral colour
dielectric (CCD) model respectively. We find that the phase
transition from hadron to quark matter is possible in a narrow
range of the parameters of nonlinear Walecka and CCD models. The
transition is strong or weak first order depending on the
parameters used. The EOS thus obtained, is used to study
the properties of hybrid stars. We find that
the calculated hybrid star properties are similar to those
of pure neutron stars.
\end{abstract}
\end{titlepage}
\vfil \eject
\begin{center}
\section{ {\bf Introduction}}
\end{center}
\par
The first model of neutron stars was proposed by Tolman, Oppenheimer
and Volkoff \c{1} in late 30's. In this model, the properties
of neutron stars were calculated by assuming that the neutron matter
consists of non-interacting degenerate gas of neutrons. Since then, a
number of improved calculations \c{2}-\c{3a}, which use different
nuclear equation of state (EOS), have been performed to determine
the composition and properties of neutron stars. The subject of
neutron stars received a new impetus after the discovery of
pulsars \c{4,4a}. In spite of two decades of work, the EOS required in
the calculation of neutron star properties is uncertain for a
number of reasons. The EOS at large baryon densities sensitively
depends on short range nucleon - nucleon interaction and this is
one source of uncertainties. Further more, at large baryon
densities, weak decay of neutrons into hyperons may be
energetically allowed and the neutron matter at high densities
may contain hyperons. Efforts have been made to include hyperon
degrees of freedom in nuclear EOS \c{4b}-\c{5b}. However, it must
be noted that the interaction of hyperons with other baryons is
not well understood and the EOS is, therefore uncertain.
\par
The quark structure of hadrons implies that at sufficiently large
nuclear densities the nuclear matter should convert itself into
quark matter. The question of transition density and the order
of the transition of the nuclear matter to quark matter is not
yet settled. The density at which the transition occurs, is
believed to be few times nuclear matter density.The lattice
calculations indicate that for nonzero quark masses the
phase transition may be weak first order or second order \c{6}. Most
of the model calculations assume it to be first order.
Thus, for large enough mass of neutron star, its core may
consist of quark matter. In addition, if the phase transition is
first order, a part of the core may consist of a mixed phase of
quark and nuclear matter. Particularly, if, at the phase
transition, the discontinuity in the baryon density between
quark and baryon matters is large, a substantial fraction of
neutron star may be in the mixed phase. It is therefore,
interesting to investigate the properties of neutron stars
having a mixture of quark and nuclear matter.
\par
In the present work, we have studied the quark-hadron phase
transition at zero temperature using two phase model and applied
it to the calculation of neutron star properties. Ideally, one
would like to investigate this phase transition using same model
in both the phases. One such method would be the lattice QCD
calculations. However, the problem of inclusion of dynamical
fermions in lattice calculations is not solved and the
calculations of equation of state for nonzero baryon chemical
potential will not be available in near future.  Thus, a number
of calculations employing different models in two phases have
been done \c{3,3a,5}. Here, we want to present a calculation where
we have used nonlinear Walecka model\c{5,5a,5b} for nuclear
matter and chiral colour dielectric model(CCDM) for quark matter \c{7}.
\par
The nonlinear Walecka model has been used extensively in nuclear
structure calculations. It has been observed that, with
nonlinear self-interaction of the $\sigma$ field, one obtains
very good agreement with the properties of nuclei over a wide
range of the periodic table \c{8}. Thus, it seems reasonable
to use such a model to describe the hadronic phase in a region
where nuclear densities are not too large in comparison with the
nuclear matter density. The equation of state calculation in
hadronic phase includes hyperons as well as leptons. The
parameters of the nonlinear Walecka model are determined by fitting the
properties of nuclear matter. However, the coupling of $\sigma$,
$\rho$ and $\omega$ mesons to hyperons is not determined by this
procedure. We have
therefore varied these couplings within a reasonable range and
investigated the effect of these on the phase transition and the
properties of stars.
\par
The CCDM has been used earlier in baryon spectroscopy \c{9}. These
calculations have
shown that the model is able to explain the static properties of
light baryons very well. Further more, when applied to quark matter
calculation, the model yields an equation of state which is quite
similar to the one obtained in lattice calculations for zero
baryon chemical potential. In particular, it shows that the energy
density calculated in CCDM is close to the energy density of free
quarks and gluons and the pressure decreases rapidly when the temperature
is close to the transition temperature \c{7}. We therefore, think that CCDM
is a qualitative improvement over the bag model, which is often
used to calculate the equation of state in quark matter. The model
has also been used in the calculation of quark star
properties \c{10}.
\par
The results of our calculations can be briefly summarised as follows.
We find that the hadron to quark matter phase transition occurs
for a narrow range of the parameter sets of the models of the two
phases. Particularly, the phase transition does not occur for smaller
values of nuclear compressibility as well as for certain parameter
sets of the CCDM (see later). This probably indicates that certain
range of parameters of nonlinear Walecka model and/or CCDM are physically not
acceptable. The neutron stars in our model, called hybrid star,
may consist of quark core, a mixed phase region and then outer
part made up of neutrons. The width of the
mixed phase depends on the parameter set used. Usually a stiffer
EOS gives higher mass or radius for a pure neutron star. But for
a hybrid star, trend is opposite. The neutron stars in our model
is found to be consistent with the observational limits.
\par
The paper is organised as follows. In Section 2, we obtain the
equations of state for nuclear and quark matter. Section 3 is devoted
to the discussion on hadron-quark phase transition. The results for
the neutron, quark and hybrid stars structure are discussed in
Section 4 and the conclusions are presented in section 5.
\begin{center}
\section{ {\bf Equations of State}}
\end{center}
\subsection { {Hadrons}}
\par
The equation of state for hadrons is calculated in the frame
work of mean field theory using the nonlinear Walecka Lagrangian given below
\c{5,5b}.
\begin{eqnarray}
{\cal L}(x)~=~\sum_{i} \bar \psi_{i} (i\gamma^{\mu}\partial_{\mu}
- m_i +g_{\sigma i}
\sigma+ g_{\omega i} \omega_{\mu} \gamma^{\mu}-
g_{\rho i} \rho^{a}_{\mu} \gamma^{\mu} T_{a} ) \psi_{i} \nonu \\
-{1 \over {4}} \omega ^{\mu \nu} \omega_{\mu \nu} +
{1 \over {2}} m^{2}_{\omega} \omega_{\mu} \omega^{\mu}
+ {1 \over {2}} ( \partial_{\mu} \sigma \partial^{\mu} \sigma-
m^{2}_{\sigma} \sigma^{2})
- {1 \over {4}} \rho^{a}_{\mu \nu} \rho^{\mu \nu}_{a}  \nonu \\ +
{1 \over {2}} m_{\rho}^{2} \rho^{a}_{\mu} \rho^{\mu}_{a}
- {1 \over {3}} bm_{N} (g_{\sigma N} \sigma)^{3} -
{1 \over {4}} c( g_{\sigma N} \sigma )^4 +
\sum_{l}\bar\psi_{l}(i\gamma^{\mu}\partial_{\mu}-m_l)\psi_{l}
\end{eqnarray}
\noindent The Lagrangian in eq(1) above includes nucleons, $\Lambda$ and
$\Sigma^-$ hyperons (denoted by subscript $i$), electrons and muons
(denoted by subscript $l$) and $\sigma$, $\omega$ and $\rho$ mesons
(given by $\sigma$, $\omega^\mu$ and $\rho^{a,\mu}$ respectively).
The Lagrangian includes cubic and quartic self-interactions of
the $\sigma$ field. The meson fields interact with baryons through
linear coupling and the coupling constants are different for nonstrange
and strange baryons. The Lagrangian defined above (without strange
baryons) has been used
extensively in nuclear structure calculations with success. The parameters
of the model are meson-baryon coupling constants, meson masses
and the coefficients of the cubic and quartic self-interactions of
$\sigma$ meson ( $b$ and $c$ respectively). The $\omega$ and $\rho$ masses
are chosen to be their physical masses.
\par
The equation of state is obtained by adopting mean field ansatz. Thus
in presence of baryons, the mesons develop nonzero vacuum expectation
values ($\bar \sigma$, $\bar \omega$ and $\bar \rho^a$ respectively).
Assuming that the baryon densities are uniform, one finds that
the time components of $\bar \omega$ and  $\bar
\rho^3$, in addition to $\bar \sigma$, are nonzero. One can then
define effective masses ($\bar m_i$)
and chemical potentials ($\bar \mu_i$) for the baryons as,
\begin{eqnarray}
\bar m_i= m_i - g_{\sigma i} \bar \sigma
\end{eqnarray}
\noindent and
\begin{eqnarray}
\bar \mu_i = \mu_i - g_{\omega i} \bar \omega - I_3 g_{\rho N} \bar \rho^{3}
\end{eqnarray}
\noindent where $I_3$ is the value of the z-component of the isospin of baryon
$i$.
The Fermi momenta ($k_i$) and number densities ($n_i$) of the baryons
are given by $k_i^2~=~{\sqrt{\bar \mu_i^2 - \bar m_i^2}}$ and
$n_i~=~{\frac {k_i^3}{3 \pi^2}}$.
For leptons, the Fermi momenta and number densities are given by
$k_l~=~\sqrt{\mu_l^2-m_e^2}$ and $n_l~=~\frac {k_l^3}{3 \pi^2}$.
\par
The condition of equilibrium under weak interactions
(assuming that neutrinos are not degenerate) give the following
relations between baryon and lepton chemical potentials
\begin{eqnarray}
\mu_{p}= \mu_{n} - \mu_{e}, ~~~~\mu_{\Lambda}= \mu_{n}, \nonu \\
\mu_{\Sigma^{-}}= \mu_{n} + \mu_{p}, ~~~~\mu_{\mu}= \mu_{e}
\end{eqnarray}
\noindent and charge neutrality gives,
\begin{eqnarray}
n_{p}= n_{e} + n_{\mu} + n_{\Sigma}
\end{eqnarray}
\noindent In addition, $n_B = n_n + n_p + n_{\Lambda} + n_{\Sigma}$ is the
total baryon density and $\mu_B = \mu_n$ is defined as the
baryon chemical potential.
\par
The mean field values of $\bar \sigma$, $\bar \omega_{0}$ and
$\bar \rho^{3}_{0}$ are determined by minimizing the energy at fixed
baryon density. Then the EOS is calculated using the expression
for pressure $P$ and energy density $E$ as given below.
\begin{eqnarray}
P~=~ {1 \over {2}} m_{\omega}^{2} \bar \omega_{0}^{2} + {1 \over
{2}} m_{\rho}^{2}(\bar \rho_{0}^{3})^2 - {1 \over {2}}
m_{\sigma}^{2} \bar \sigma^{2} - {1 \over {3}} bm_{N}(g_{\sigma
N} \bar \sigma)^3 - {1 \over {4}}c(g_{\sigma N} \bar \sigma)^4
\nonu \\
+ \sum_{i} P_{FG}(\bar m_{i}, \bar \mu_{i}) +\sum_{l} P_{FG}(m_{l},
\mu_{l})
\end{eqnarray}
\begin{eqnarray}
E~=~ {1 \over {2}} m_{\omega}^{2} \bar \omega_{0}^{2} + {1 \over
{2}} m_{\rho}^{2}(\bar \rho_{0}^{3})^2 + {1 \over {2}}
m_{\sigma}^{2} \bar \sigma^{2} + {1 \over {3}} bm_{N}(g_{\sigma
N} \bar \sigma)^3 + {1 \over {4}}c(g_{\sigma N} \bar \sigma)^4
\nonu \\
+\sum_{i} E_{FG}(\bar m_{i}, \bar \mu_{i}) + \sum_{l} E_{FG}(m_{l},
\mu_{l})
\end{eqnarray}
\noindent In the above equations $P_{FG}$ and $E_{FG}$ are the
relativistic non-interacting pressure and energy density of the
fermions. Nonlinear Walecka model has eight parameters out of
which five are determined
by the properties of nuclear matter. These are nucleon
coupling to scalar (${g_{\sigma} \over {m_{\sigma}}}$), vector
(${g_{\rho} \over {m_{\rho}}}$) and isovector mesons (${g_{\omega} \over
{m_{\omega}}}$) and  the two coefficients in scalar
self interaction i.e. $b$ and $c$ . These are obtained by fitting
saturation values of the binding energy/nucleon ($-16
MeV$ ), baryon density ($0.15fm^{-3}$ ), symmetry energy
coefficient ($32.5 MeV$ ), Landau mass ($0.83 m_N$).
The nuclear compressibility is somewhat
uncertain and therefore, we have varied it between $250MeV-
350MeV$. The values of these parameters along with
compressibility are presented in Table 1.
\par
The other three parameters are coupling constants of
hyperon-meson interaction and are not well
known. These cannot be determined from nuclear matter properties
since the nuclear matter does not contain hyperons. Further
more, properties of hypernuclei do not fix these parameters in
a unique way. In the literature, a number of choices have been
made. These along with our choice are listed below.
\begin{enumerate}
\item choose them to be same as nucleon-meson coupling
constants (Universal coupling ) \c{5},
\item choose them to be $\sqrt{2/3}$ times nucleon-meson coupling
constants \c{5a},
\item choose them to be $1/3$ times nucleon-meson coupling
constants \c{5b}
\item It is claimed that the second choice follows from quark counting
rule. However, using SU(6) quark wavefunctions of
baryons and assuming that the mesons do not couple to strange
quarks, we find that the ratio of meson-hyperon and
meson-nucleon couplings are $2/3$, $2/3$ and $1$ for $\sigma$,
$\omega$ and $\rho$ mesons respectively.
\end{enumerate}
\par
In our calculation, we have used the above couplings to
study the nuclear EOS and star characteristics.
\subsection{{Quarks}}
\par
The Colour Dielectric Model (CDM) is based on the idea of Nelson
and Patkos \c{11}. In this model, one  generates the confinement of
quarks and gluons
dynamically through the interaction of these fields with scalar
field. In the present work, we have used the chiral
extension \c{7} (CCDM) of this model to calculate quark
matter EOS. The Lagrangian density of CCDM is given by
\begin{eqnarray}
{\cal L}(x)= \bar\psi(x)\big \{ i\gamma^{\mu}\partial_{\mu}-
(m_{0}+m/\chi(x) U_{5}) + (1/2) g
\gamma_{\mu}\lambda_{a}A^{a}_{\mu}(x)\big \}\psi \nonumber \\
+f^{2}_{\pi}/4 Tr ( \partial_{\mu}U
\partial^{\mu}U \dagger ) -
1/2m^{2}_{\phi} \phi^{2}(x) \nonumber \\ -(1/4)
\chi^{4}(x)(F^{a}_{\mu\nu}(x))^{2} + (1/2)
\sigma^{2}_{v}(\partial_{\mu}\chi(x))^{2} -U(\chi)
\end{eqnarray}
\noindent where $U = e^{i\lambda_{a}\phi^{a}/f_\pi}$ and $U_{5} =
e^{i\lambda_{a}\phi^{a}\gamma_{5}/f_{\pi}}$, $\psi(x)$,
$A_{\mu}(x)$, $\chi(x)$ and $\phi(x)$ are quark, gluon, scalar (
colour dielectric ) and meson fields respectively, $ m_{\phi}$
and m are the meson and quark masses, $ f_{\pi}$ is the pion
decay constant, $F_{\mu\nu}(x)$ is the usual colour
electromagnetic field tensor, g is the colour coupling constant
and $\lambda_{a}$ are the Gell-Mann matrices.  The flavour
symmetry breaking is incorporated in the Lagrangian through the
quark mass term $(m_{0}+m/\chi U_{5})$, with $m_{0} = 0$ for $u$
and $d$ quarks. So the masses of $u$, $d$ and $s$ quarks are
$m$, $m$ and $m_{0} + m$ respectively. The self interaction
$U(\chi)$ of the scalar field is assumed to be of the form
\begin{eqnarray}
\alpha B
\chi^2(x)[
1-2(1-2/\alpha)\chi(x)+(1-3/\alpha)\chi^2(x)]
\end{eqnarray}
\noindent so that $U(\chi)$ has an absolute minimum at $\chi =
0$ and a secondary
minimum at $\chi = 1$. The interaction of the scalar field with quark
and gluon fields is such that quarks and gluons can not exist in the
region where $\chi= 0$.
In  the  limit  of  vanishing  meson  mass,  the
Lagrangian of eqn.(8) is invariant under chiral transformations of quark
and meson fields.
\par
The calculation of equation of state proceeds as follows. We
assume that, in presence of nonzero quark/anti-quark densities,
the square of meson fields may develop nonzero vacuum
expectation values ($<\phi^2>$) \c{7,10}. This assumption is
analogous to the assumption that, in linear $\sigma$ model \c{7a}, the
$\sigma$ field acquires nonzero vacuum expectation. In
\c{7}, it has been shown that this occurs when the quark
density exceeds certain critical value and one of the effects of
nonzero $<\phi^2>$ is that the effective quark masses decrease
with the increase in $<\phi^2>$. Thus at large quark densities,
one obtains an equation of state similar to the equation of
state of free quarks and gluons. Further more, we adopt mean
field approximation to calculate the colour dielectric field
($\chi$) in quark matter. With these assumptions, the
thermodynamic potential for the quark matter is given by,
\begin{eqnarray}
\Omega~=~{1 \over {4\pi^2}} \sum_{i} \big \{\big
[\mu_{i}k_{i}(\mu_{i}^{2} - {5\over {2}} {m_{i}^{*}}^{2})~+~
{3\over {2}}ln\big ({{\mu_{i}~+~k_{i}}\over {m_{i}^{*}}})]  \nonu \\
-~{\alpha_{s} \over {\pi}}\big [{3\over {2}}\big (\mu_{i}k_{i} -
{{m_{i}^*}^{2}} ln\big ({{\mu_{i}~+~k_{i}} \over {{m_{i}}^{*}}}))^2 -
k_{i}^4]\}
\end{eqnarray}
where $i= u, d, s$. Also, $k_{i}~=~\sqrt{\mu_{i}^{2}~-~{{m_{i}^{*}}^{2}}}$
and this becomes equal to Fermi momentum for $\alpha_{s}~=~0$.
Here $\Omega$ is calculated upto second order in quark-gluon
interaction.
\par
In addition, chemical equilibrium under weak decay and charge
neutrality imply,
\begin{eqnarray}
\mu_{d}  = \mu_{u}+ \mu_{e} \nonu \\
\mu_{s}  = \mu_{u} + \mu_{e}
\end{eqnarray}
and
\begin{eqnarray}
2/3 n_{u} - 1/3 n_{d} -1/3 n_{s} - n_{e}  = 0
\end{eqnarray}
The baryon density $n_{B}=1/3\sum_{i}(n_{i})$ where $i= u,d,s$ and
baryon chemical potential is defined as
$\mu_{B}~=~\mu_{d}~+~\mu_{u}~+~\mu_{s}$.
\par
The mean field
values of $\chi$ and $F_{\phi}$ are calculated by minimising
$\Omega$. Equivalently, one can solve the equations of motion for
$\chi$, $F_{\pi}$, $F_{K}$ and $F_{\eta}$ as obtained from the
Lagrangian. We find that $<\vec\pi^{2}>$ alone develops non zero
value in the quark matter. $<K^2>$ and $<\eta^2>$ remain zero
throughout the range of densities considered. This implies that
strange quark mass remain constant in the medium. On the other
hand, $u$ and $d$ quark masses change in the medium.
The pressure $P$ and energy density $E$ of the quark matter are
calculated using the relations, $P~=~- \Omega$ and
$E~=~\Omega~+~\mu_{i} n_{i}$. Here, the number density of the
quark of $i$th type is given by $n_{i}~=~-
(\partial\Omega/\partial\mu_{i})$
\par
The parameters of the CCDM are $u$ and $d$ quark mass,
strange quark mass, strong coupling constant $\alpha_{s}$, bag
pressure $B$ and $\alpha$. These are obtained by fitting the baryon
masses. Earlier calculations \c{9} show that these are not
determined uniquely by the fitting procedure. In particular, it
has been found that good fits to baryon masses are obtained for
$0.6 GeV\le m_{GB} \le 3GeV$ , $m_{q}(u,d) \le 125MeV$, $m_q(s) \sim
300MeV$ and $B^{1/4}\le 150MeV$, where glueball mass $m_{GB}$ is
defined as $m_{GB}^{2}= 2B\alpha/\sigma_{v}^{2}$. The fits are
better for lower values of $m_{GB}$, $m_{q}(u,d)$ and $B$. Since
we do not consider scalar field excitations, the quark matter
EOS does not depend on $m_{GB}$.
\begin{center}
\section{{\bf Hadron - Quark Phase Transition} }
\end{center}
\par
Having determined the equations of state of neutron and quark
matter, the phase transition point is determined by adopting
Gibb's criterion. That is, the point at which the free energies
( or pressure ), for  given chemical potential, are equal gives
the phase transition point. In the present context, the baryon
and electron chemical potentials ( $\mu_B$ and $\mu_e$
respectively ) are the two independent chemical potentials.
However, if we assume the charge neutrality,
$\mu_e$ is no more independent. Then the crossing of the
pressure curves for two phases in $P$ -$\mu_B$ plane gives the
phase transition point. Glendenning \c{5a}, on the other hand
considered a case where, in the mixed phase, neutron and quark
matters are not charge neutral but the mixture as a whole is.
This aspect has been studied further by Heiselberg et.al.
in ref \c{12}. We are not considering this situation here.
\par
For a first order transition, the derivatives of $P$- $\mu_B$
curve for the two phases at the phase transition point are not
equal and the difference in the two derivatives gives the
discontinuity in the density of the two phases at the transition
point. The two phases coexist in this range of density. The
latent heat of transition is given by the difference in the energy
densities of the two phases at the critical point. As mentioned
earlier, the lattice QCD calculations indicate that hadron -
quark phase transition may be weakly first order or second
order. In a calculation, such as ours, we will necessarily have
a first order transition since two different models are employed
to calculate the properties of quark and hadron phases. However,
our calculation shows that the phase transition can be made
weaker by varying the compressibility of the nuclear matter as
well as the meson- strange baryon coupling within reasonable
limits. For example, Table 2 shows that the latent heat can be
reduced from $154 MeV/fm^3$ to $96 MeV/fm^3$ by changing
the compressibility from $250 MeV$ to $300 MeV$. Thus, it is
possible to mimick the second order phase transition in this
manner.
\par
Our calculation shows that as the nuclear compressibility is
increased from $250 MeV$, the chemical potential ( and baryon
density ) at which the phase transition occurs decreases. ( see
Fig. 1) This can be understood as follows. For a given
parametrization, the nucleon- nucleon repulsive interaction increases with
compressibility and hence the slope of the $P- \mu_{B}$ curve at the
transition point i.e. the baryon density decreases. The
dependence of latent heat on compressibility is given in Table 2.
\par
As mentioned earlier, the strange quark - meson coupling
constant in hadron phase is not well known. We have therefore,
varied this coupling constant as mentioned in earlier section
and investigated its effect on the properties of phase
transition. We find that as the coupling constant is reduced
from the universal coupling (
$g_{H\alpha}/g_{N\alpha}$=1, $\alpha=\sigma,~\omega$ or $\rho$ and
$H=\Lambda$ or $\Sigma^{-}$ ) the
slope of the $P- \mu_B$ curve ( Fig.2 - 5) increase which
leads to the
increase in transition density and as well as reduction in the
latent heat (Fig.6 and Table 2). The EOS becomes softer with
decrease in hyperon
couplings. This can be understood as follows. With decrease in
hyperon couplings, it is energetically favourable to convert
nucleons into hyperons as hyperons do not feel the predominantly
repulsive force. As a result with decreasing coupling more and
more hyperons get populated thereby reducing the energy further.
\par
As mentioned in the previous section, the quark matter EOS has
been obtained by using the parameter sets which fit the baryon
masses. We find that the transition density for hadron quark
phase transition decreases rapidly with the decrease in $B$. For
$B^{1/4} \le 140MeV$, the transition density is smaller than the
nuclear matter density. This is unphysical, since it implies
that we should have quark matter at nuclear matter densities.
This happens because the contribution of the dielectric field to
the pressure, which is negative, is proportional to $B$. So for
smaller $B$, the pressure of the quark matter increase and this
makes the quark matter stable at lower chemical potentials and
densities. This type of behaviour has also been noticed for the
bag model EOS where one requires $B^{1/4} \sim 150MeV$ \c{3a}. At
this point, we would like to note that $B$ can not be increased
arbitrarily if one insists on a reasonable fit to baryon masses.
Thus in CCDM, a very restricted parameter sets give reasonable
values for transition densities. One such set ($B^{1/4}= 152.1
MeV$, $m_{q(u,d)}= 91.6 MeV$, $m_{q(s)}= 294.9 MeV$ $\alpha= 36$
and strong coupling $\alpha_s= 0.08$) has been used in the
calculations reported here.
\par
Finally, let us consider the change in strangeness fraction at
the transition point. The strangeness fraction in quark and
nuclear matter are defined by the ratio strange quark density/
baryon density and strange hadron density/baryon density
respectively. Note that for equal mass quarks
and $\Lambda$ matter this ratio is unity. We find that generally
at the transition point the strangeness fraction is larger in
quark matter. This is shown in Fig.7, where strange fraction is
plotted as a function of baryon density. Note that in the
coexistence region, the system consist of a mixture of quark and
nuclear matter. The strangeness fractions in the mixed phase are
calculated using the linear relation
$f_s(M)~=~(1-\chi)f_s(H)~+~\chi f_s(Q)$, where $\chi$ is the
concentration of the quark matter in the mixed phase and $f_s$ is the
strangeness fraction, $M$, $H$, and $Q$ denoting the mixed,
hadronic and quark phase respectively. Similar relation is used
to calculate the corresponding baryon density as well. It is
evident from the graph that for a large mixed phase region, the
jump in the strangeness fraction from hadronic to quark phase is
larger.
\par
The presence of considerable amount of strangeness in the
hadronic sector in the form of hyperons has important
consequences on the mechanism of phase transition and core of
the neutron stars. Some of the
possible mechanisms are discussed in ref[ Alcock, Farhi and Olinto \c{13}].
The conversion via two flavour has been considered in ref \c{14},
where as in ref \c{15}, Olinto has started with the assumption that
strange matter has been seeded into neutron star from outside.
On the other hand, presence of considerable amount of hyperons
near the transition point suggest that inside a cold neutron
star conversion from neutron to strange quark matter may occur
through the clustering of $\Lambda$'s \c{13}. For a hot neutron star
the conversion to strange matter will occur due to thermal $\Lambda$'s.
Also, along with these processes, there will be usual
strangeness changing weak decay to convert excess $d$ quarks to
$s$ quarks. We believe that these points need further
investigation.
\begin{center}
\section{{\bf Neutron star models} }
\end{center}
\par
Quark cores can exist inside a neutron/hybrid star only for a narrow range of
parameter sets in our model. The extent of quark core will be
higher for larger compressibility and larger hyperon couplings.
\par
The structure of a neutron star is characterised by its mass
and radius. Additional parameters of interest are the moment of
inertia, the surface red shift $z$ and the relativistic
Keplarian rotation period $P_K$ defined as \c{3}:
\begin{equation}
P_K = 0.026\sqrt{(R/km)^{3}\over{(M/M_{\odot})}} [ms]
\end{equation}
\noindent as a function of the central density $\rho_c$ of the star. These
are important for the dynamics and transport properties of
pulsars.
\par
The equations that describe the hydrostatic equilibrium of
degenerate stars without rotation in general relativity is called
Tolman-Oppenheimer-Volkoff (TOV) equations, which is given in ref
\c{2a,10,16,17}. These equations can be numerically integrated, for a
given central density, to obtain the radius $R$ and the
gravitational mass $M$ of the star. The moment of inertia
$I$ of the rotating neutron star, is also calculated [see ref
\c{2a,17}] for the corresponding central density. To integrate the TOV
equations, one needs to know the equation of state $P(\rho)$,
for the entire expected density range of neutron star, starting
from the higher density at the center to the surface density.
The composite equation of state for the entire neutron star
density span, was constructed by joining the nonlinear Walecka
hadronic equation of state (eqn.6-7) to that given by (a)
Negle and Vautherin \c{18} for
the density range $10^{14}$ to $5 \times 10^{10}$
$gm/cm^3$, (b) Baym, Pethick and Sutherland \c{19} for the range $5
\times 10^{10}$ to $10^3$ $gm/cm^3$ and (c) Feynman, Metropolis and
Teller \c{20} for densities less than $10^3 gm/cm^3$.
\par
The results for star structure parameters are listed in Table 3.
Fig.8, and Fig.9  show plots of mass vs central density and
mass vs moment of inertia respectively. We have plotted the curves for
hadronic EOS with hyperon couplings (1), (2) and (4) as given in
section 2.1 and fixed compressibility $300MeV$. In case of quark matter,
we have used the interacting CCDM with the parameter set discussed
in section 3.
For neutron stars a stiffer EOS results in a
larger maximum mass and radius \c{3,5b}. So for heavier stars, one needs
larger compression constant $K$ or larger hyperon couplings.
Similarly for quark stars, one needs larger $B$ or $\alpha_s$.
This dependence is reversed for hybrid stars where a more
repulsive interaction yields lighter
stars as can be seen in Fig.8 and Table 3. This is because a
stiffer EOS implies lower critical baryon densities in the
hadronic sector and hence larger quark cores. Since quark EOS
yield smaller maximum masses, this reduces the
maximum mass of the hybrid stars (Fig.8 and Table 3).
\par
Here we would like to mention that the most accurately
determined mass is that of PSR1913+ 16 with
$M/M_{\odot}= 1.44\pm 0.003$ \c{4a}. The observational lower limit on
the moment of inertia \c{21} is $I= 40 M_{\odot} km^2 \simeq 8\times
10^{44} gm/cm^3$. The red shift, which is not measured for any
neutron star with known mass, seem to lie in the range 0.2 -
0.5. In principle, the above observational results on neutron
stars should put some constraint on the EOS. But in practice,
most of the EOS reproduce consistent results as found by other
authors as well \c{3,5b}. This can also be seen from our results for
characteristics of the maximum mass as given in Table 3.
Table 3 contains the results for neutron stars models $HM~I$,
$HM~II$ and $HM~III$ corresponding to three hyperon couplings
((1), (2) and (4), see section 2.1) with compressibility
$300MeV$ and for hybrid stars models $Hybrid~ I$, $Hybrid~ II$ and
$Hybrid~ III$ respectively.
In our model, the maximum gravitational mass for stable non-rotating
neutron stars are in the range $1.86M_{\odot}- 1.6M_{\odot}$. This
decrease in maximum mass, is due to softness of EOS. That is
because of decrease in hyperon couplings as explained earlier.
The corresponding radius increases from $10.69km- 11.00km$,
whereas, red shift and moment of inertia decrease from $0.44-
0.32$ and $1.76\times 10^{45}~g~ cm^2- 1.52\times 10^{45}~g~ cm^2$
respectively.
Similarly, the hybrid star maximum mass corresponding to the
three hyperon couplings are in the range $1.47M_{\odot}-
1.49M_{\odot}$. The corresponding variation of radius, red shift
and moment of inertia are $12.36km- 12.0km$, $0.242- 0.256$ and
$1.75\times 10^{45}~ g~ cm^2- 1.71\times 10^{45}~g~ cm^2$. The small
variation of hybrid star properties implies that hadronic EOS
does not have strong influence on the masses and radii of the hybrid
star as observed by other authors as well \c{3}. We have also calculated
the Keplarian rotation period. The minimal rotation periods for neutron
stars are in the range $0.66ms- 0.75ms$. The corresponding range for
hybrid stars is $0.93ms- 0.89ms$.
So the periods for both neutron and hybrid stars, in our model,
are comparable with the limits obtained by other authors \c{3}.
Thus, the general characteristics of stable star, in our model,
are compatible with the observational estimates.
\par
The post glitch data set from the vela pulsar indicates that crust
superfluid comprises of about $3.4 \times 10^{-2}$ of the stars
moment of inertia \c{22}. This is lower bound on the fractional
moment of inertia of the entire crust. We have calculated
$\alpha~=~I_p/I$, where $I_p$ is the moment of inertia of the
pinned superfluid region and $I$ is the total moment of inertia
of the star. This region is defined as the radial
extent corresponding to the density $2\times 10^{14}~-~2\times
10^{13}$ $gm/cm^3$. Table 3 shows that the value of $\alpha$
is sensitive to the values of hyperon couplings.
The purpose of
this calculation is to compare an observational feature of pulsar
glitches with predicted theoretical values of neutron and hybrid
stars in the present model.
\begin{center}
\section{{\bf Summary }}
\end{center}
\par
We have described the hadron - quark phase transition at zero
temperature in the frame work of relativistic mean field theory
for hadronic sector and chiral Colour Dielectric model for quark
sector. We have studied the hadronic EOS and phase transition
by varying the compressibility from $250MeV- 350MeV$.
Using a simple quark counting rule, we get a hyperon coupling
which is different compared to those used by other authors \c{5,5a,5b}. We
have given results for two other couplings found in the
literature as well. We find that critical $\mu_B$ and critical $n_B$ can
be reduced by increasing the hyperon coupling or
compressibility. Moreover, latent heat can also be reduced by
decreasing the hyperon coupling or varying the compressibility.
This further implies a weaker first order transition. In quark
sector, we have taken the parameter sets from baryon
spectroscopy. We find that
only for a very narrow range of parameters in CCDM, one can get
phase transition. In CCDM, the quark matter EOS depends
sensitively on quark masses, strong coupling and bag pressure.
For $B^{1/4} < 140MeV$ the transition density is less than the
nuclear matter density which is unphysical. Here we would like
to mention that in
our calculation, it is not possible to change any of the parameters
arbitrarily, as it would destroy the fit of baryon masses.
\par
We find that a considerable amount of strangeness is present in
the hadronic phase at the transition point. Also there
is a large change in the strangeness fraction
from hadronic to quark phase. Several authors \c{14} have studied the
production rates of strangeness during the phase transition from
nuclear to quark matter. But as predicted by Alcock et.al. \c{13}, the
rate of transition to quark matter may be much faster, if
strangeness is already present in the hadronic phase. It will be
interesting to study the mechanism of transition to quark phase
and the rate of strangeness production in such circumstances.
\par
The neutron star characteristics are calculated for different
EOS obtained from Walecka models and CCDM. It is clear that the
mass limits for stars are sensitive to the parameter sets. We
find that CCDM gives softer EOS compared to bag models and hence
lower masses and radii. Hybrid star masses are lower compared to pure
neutron stars because of the quark cores.
We found that the stars properties are not much sensitive to the
composition of stars. Also these characteristics obey the
observationally inferred limits. So, it is difficult to
distinguish between the stars of different composition
observationally. Hence the most massive pulsars observed
so far could possibly either a neutron star or a hybrid star.
We do not get the layered structure as
predicted in reference \c{23}. In fact, we find that parameter sets
which predict a stable quark phase at lower densities, also
predict a unstable isospin symmetric nuclear matter at lower
densities. Here we would like to point out that the nuclear to
quark matter phase transition can
happen through some intermediate phases as well. The nuclear
matter may go over to a pion or kaon \c{24} condensate phase
through a second order transition which then goes over to quark
phase at higher densities. In such case the transition to quark
phase happens at much higher densities. Some other authors have
tried to explore the possibility of second order transition \c{5b},
which we do not consider here.
\vfil
\eject

\vfill
\eject
\newpage
\begin{table}
\vspace{0.2in}
\caption { Coupling constants for several compresibility $K$ and
$B/A$=-16 $MeV$, $\rho=0.15~fm^{-3}$, $a_{sym}=32~MeV$ and
${m^{*}}_{sat}/m=0.8$}
\begin{center}
\begin{tabular}{|c|c|c|c|c|c|}
\hline
\multicolumn{1}{|c|}{$K$} &
\multicolumn{1}{|c|}{$(g_s/m_s)^2$} &
\multicolumn{1}{|c|}{$(g_w/m_w)^2$} &
\multicolumn{1}{|c|}{$(g_{\rho}/m_{\rho})^2$} &
\multicolumn{1}{|c|}{$b$} &
\multicolumn{1}{|c|}{$c$} \\
\multicolumn{1}{|c|}{($MeV$)} &
\multicolumn{1}{|c|}{($fm^2$)} &
\multicolumn{1}{|c|}{($fm^2$)} &
\multicolumn{1}{|c|}{($fm^2$)} &
\multicolumn{1}{|c|}{} &
\multicolumn{1}{|c|}{} \\
\hline
250&9.216&4.356&5.025&0.008209&0.007385\\
300&8.492&4.356&5.025&0.002084&0.02780\\
350&7.820&4.356&5.025&-0.004618&0.05015\\
\hline
\end{tabular}
\end{center}
\end{table}
\vfill
\eject
\newpage
\begin{table}
\caption{Characteristics at the critical point. The columns
correspond to compressibility ($K$), hyperon couplings ($HC$),
critical pressure ($P_c$), critical chemical potential
($\mu_c$), energy of the hadronic phase at critical point
($E_{cH}$), latent heat ($L$), baryon density in the hadronic
phase at critical point ($n_{B(cH)}$) and baryon density width
of the mixed phase ($\Delta n_B$) }
\hskip 0.5 in
\begin{center}
\begin{tabular}{|c|c|c|c|c|c|c|c|}
\hline
\multicolumn{1}{|c|}{$K$} &
\multicolumn{1}{|c|}{$HC$} &
\multicolumn{1}{|c|}{$P_c$} &
\multicolumn{1}{|c|}{$\mu_c$} &
\multicolumn{1}{|c|}{$E_{cH}$} &
\multicolumn{1}{|c|}{$L$} &
\multicolumn{1}{|c|}{$n_{B(cH)}$} &
\multicolumn{1}{|c|}{$\Delta n_{B}$} \\
\multicolumn{1}{|c|}{($MeV$)} &
\multicolumn{1}{|c|}{} &
\multicolumn{1}{|c|}{$MeV/fm^3$} &
\multicolumn{1}{|c|}{$MeV$} &
\multicolumn{1}{|c|}{($MeV/fm^3$)} &
\multicolumn{1}{|c|}{$MeV/fm^3$} &
\multicolumn{1}{|c|}{$fm^{-3}$} &
\multicolumn{1}{|c|}{$fm^{-3}$} \\
\hline
&(1)&62.2&1179.5&490.7
&123.2&0.47&0.11 \\
300.&(2)&78.5&1200.0&581.2
&96.4&0.53&0.09 \\
&(4)&75.0&1194.0&603.7
&58.6&0.55&0.06 \\
\hline
250.&(2)&269.3&1428.0&1236.6
&153.6&1.0&0.11 \\
\hline
350.&(2)&58.0&1163.5&480.8
&128.5&0.45&0.11 \\
\hline
\end{tabular}
\end{center}
\end{table}
\vfill
\eject
\newpage
\begin{table}
\caption { Star characteristics for different models. Here
$IQM$ for interacting quark matter and $HM$ for hadronic matter.}
\hskip 0.5 in
\begin{center}
\begin{tabular}{|c|c|c|c|c|c|c|c|}
\hline
\multicolumn{1}{|c|}{$\rho_c$} &
\multicolumn{1}{|c|}{$R$} &
\multicolumn{1}{|c|}{$M/M_{\odot}$} &
\multicolumn{1}{|c|}{$z$} &
\multicolumn{1}{|c|}{$I$} &
\multicolumn{1}{|c|}{$P_{K}$} &
\multicolumn{1}{|c|}{$I_{p}/I$} &
\multicolumn{1}{|c|}{$Model$} \\
\multicolumn{1}{|c|}{($g~cm^{-3}$)} &
\multicolumn{1}{|c|}{($km$)} &
\multicolumn{1}{|c|}{} &
\multicolumn{1}{|c|}{} &
\multicolumn{1}{|c|}{($g~cm^{2}$)} &
\multicolumn{1}{|c|}{$ms$} &
\multicolumn{1}{|c|}{} &
\multicolumn{1}{|c|}{} \\
\hline
3.60$\times 10^{15}$&8.14&1.38&0.41&8.22$\times
10^{44}$&0.52&--&$IQM$ \\
\hline
2.50$\times 10^{15}$&10.69&1.86&0.44&1.76$\times
10^{45}$&0.66&2.98$\times 10^{-2}$&$HM~ I$ \\
2.50$\times 10^{15}$&10.93&1.72&0.37&1.52$\times
10^{45}$&0.72&6.10$\times 10^{-2}$&$HM~ II$ \\
2.00$\times 10^{15}$&11.00&1.60&0.32&1.52$\times
10^{45}$&0.75&2.68$\times 10^{-2}$&$HM~ III$ \\
\hline
0.84$\times 10^{15}$&12.36&1.47&0.24&1.75$\times
10^{45}$&0.93&2.94$\times 10^{-2}$&$Hybrid~ I$ \\
&&&&&&& ($IQM$+$HM~ I$)\\
0.99$\times 10^{15}$&12.94&1.49&0.23&1.72$\times
10^{45}$&0.99&5.4$\times 10^{-2}$&$Hybrid~ II$ \\
&&&&&&& ($IQM$+$HM~ II$)\\
1.03$\times 10^{15}$&12.00&1.49&0.26&1.71$\times
10^{45}$&0.89&2.39$\times 10^{-2}$&$Hybrid~ III$\\
&&&&&&& ($IQM$+$HM~ III$)\\
\hline
\end{tabular}
\end{center}
\end{table}
\vfill
\eject
\newpage
\begin{figure}
\centerline {FIGURE CAPTIONS}
\caption {Pressure vs. chemical potential {\bf (a)}
Interacting quark matter ($Q.M.$), {\bf (b)} Hadronic matter ($H.M.$), $K$=
250$MeV$
, hyperon coupling ($HC$) (2), {\bf (c)} $H.M.$, $K$= 300$MeV$, $HC$ (2),
{\bf (d)} $H.M.$, $K$= 350$MeV$, $HC$ (2) and {\bf (e)} Bag model with
bag constant $B^{1/4}$= 152.1$MeV$}
\caption {Pressure vs. chemical potential {\bf (a)} $Q.M.$
{\bf (b)} $H.M.$, $K$= 300$MeV$, $HC$ (1)}
\caption {Pressure vs. chemical potential {\bf (a)} $Q.M.$
{\bf (b)} $H.M.$, $K$= 300$MeV$, $HC$ (2)}
\caption {Pressure vs. chemical potential {\bf (a)} $Q.M.$
{\bf (b)} $H.M.$, $K$= 300$MeV$, $HC$ (4)}
\caption {Pressure vs. chemical potential {\bf (a)} $Q.M.$
{\bf (b)} $H.M.$, $K$= 300$MeV$, $HC$= 0.75  }
\caption {EOS with first order phase transition from $H.M.$ to
$Q.M.$ with compressibility 300$MeV$ {\bf (a)} $HC$ (1), {\bf (b)}
$HC$ (2) and {\bf (c)} $HC$ (4)}
\caption {Variation of strangeness fraction with baryon density
for the EOS in Fig.6, {\bf (a)} $HC$ (1), {\bf (b)} $HC$ (2) and
{\bf (c)} $HC$ (4)}
\vfil
\eject
\newpage
\caption {Mass vs. central density of neutron ($K$= 300$MeV$),
quark and hybrid stars {\bf (a)} neutron star (N.S.), $HC$ (1), {\bf (b)} N.S.,
$HC$ (2), {\bf (c)} N.S., $HC$ (4), {\bf (d)} quark star .
Corresponding curves for hybrid star are denoted by $Hybrid~ I$,
$Hybrid~ II$ and $Hybrid~ III$ respectively. }
\caption {Moment of inertia vs. mass of neutron ($K$= 300$MeV$),
quark and hybrid stars {\bf (a)} N.S., $HC$ (1), {\bf (b)} N.S.,
$HC$ (2), {\bf (c)} N.S., $HC$ (4), {\bf (d)} quark star .
Corresponding curves for hybrid star are denoted by $Hybrid~ I$,
$Hybrid~ II$ and $Hybrid~ III$ respectively.}
\end{figure}
\vfill
\eject
\end{document}